\documentclass[reqno]{amsart} 
\usepackage{amssymb,amsmath,amsfonts,amsthm,comment,mathrsfs,times,graphicx}
\usepackage[T1]{fontenc}
\usepackage{lmodern}

\usepackage{geometry}                
\title[A theory...composite F-actin and vimentin networks]{A theory to describe emergent properties of composite F-actin and vimentin networks}
\author[H. Lopez-Menendez \& L. Gonzalez-Torres]{Horacio Lopez-Menendez$^1$ \and Libardo Gonzalez-Torres$^2$}
\date{\today
\\
    $^1$Institut Jacques Monod (IJM), CNRS UMR 7592 et Universit\'e Paris Diderot, 75013 Paris, France.\\ To whom correspondence should be addressed. E-mail:\textbf{ horacio.lopez.menendez$@$gmail.com}\\%
    $^2$Institute of Science and Technology, Federal University of the Valleys of Jequitinhonha and Mucuri, 39100-000, Minas Gerais, Brazil.\\[2ex]%
}

\begin{document}
\begin{abstract}
The synthetic biopolymeric gels demand a great interest as bio-materials to mimic many biological scaffolding structures, which can contribute to a better understanding of the cytoskeleton-like structural building blocks and soft nanotechnology. In particular semiflexible F-actin and vimentin intermediate filaments (IF) form complex networks, and are key regulators of cellular stiffness. While the mechanics of F-actin networks or IF have already been characterised, the interaction between this two networks is largely unknown. Experimental studies using large deformations rheology show that co-polymerisation of F-actin and IF can produce composite networks either stronger or weaker than pure F-actin networks. We theoretically verify these effects developing a model into the framework of nonlinear continuum mechanics, in which we define a free energy functional considering the role of the entropic-elastic for semiflexible networks with transient crosslinks and also an energetic term to describe the interaction parameter which allows the coupling among the two networks. We validate the theoretical model with measurements performed performed by Jensen et al. on large deformations rheological experiments with different concentrations of actin and vimentin 
\end{abstract}

\maketitle


\section{Introduction}

The mechanical scaffolding of the cell, called cytoskeleton, is defined by bio-polymeric structures such as F-actin, microtubules, and intermediate filaments (IF) that creates networks which are critical in determining the mechanical properties of the cells. Thus, the cytoskeleton conducts many mechanical duties such as mechano-sensing, motility, contraction, division and extrusion. Their dysfunctions are strongly associated with several pathological conditions. In vivo, it can be found from dense amorphous networks to well-organized bundled arrays. These varieties of assemblies are very dynamic, and evolving by non-equilibrium actin polymerization/depolymerization and also by the active force such as myosin motors. An ideal system for such studies is the in vitro network, as it provides a well-controlled environment. Previous in vitro studies have reported the mechanics of either single filaments \cite{gittes1993, mucke2004}, or networks of filaments comprised of single biopolymer species \cite{janmey1991, gardel2004, Gardel2008}. Complementing the advances developed using in-vitro networks, a large number of theoretical and computational models have been developed, providing new ways of thinking about cellular mechanics. In this sense, some microstructural approaches based on the worm-like chain model represent an excellent description of the actin mechanics \cite{broedersz2014, lopez2016, meng2017, vernerey2018, ferreira2018}. Furthermore, several computational efforts evaluating a large-scale fiber models have recently made a substantial progress, but they are still limited to passive situations without considering the internal stresses due to the effect of the entanglements and polymerisation dynamic \cite{Kim2009b, borau2012}. On the other side, in the context of hydrogels, a much more relevant work has been made on interpenetrating polymer networks (IPN) which consists of two or more polymer networks, at least one of which is polymerised and / or crosslinked in the immediate presence of the other. The polymer networks are interlaced on a molecular scale but not covalently bonded to each other. Above glass transition temperature, IPN are capable to achieve large deformations and to manifest high toughness, Mullins effect and necking instabilities \cite{nakajima2013, zhao2014, ducrot2014}. In order to improve the understanding of the micromechanics of IPN, a constitutive modelling of interpenetrating networks has also been proposed \cite{suo2009,zhao2012}.

Nevertheless, studies combining F-actin and IF are few, if we consider that together represent the majority of the intracellular network \cite{luby1999}. This sort of studies are notably interesting because the co-polymerisation of the two networks will shape a resultant structural state strongly modified by alterations in assembly kinetics and steric constraints, where the presence of an IF network is likely to alter the actin assembly \cite{pelletier2009, kayser2012}. Thus, gaining a deep understanding of the emergent behaviour of composite networks will provide better ways to control and build complex structures. In this regard Jensen et al.  elaborated a crosslinked F-actin network interpenetrated with a vimentin IF network and used bulk rheology to investigate the composite network mechanics in both the linear and nonlinear regimes. They found that co-polymerisation with vimentin strengthens F-actin networks when actin crosslinks are abundant, as expected from the overall increase in the amount of polymer in the network. Unexpectedly they found that the mechanical response of the F-actin networks are weakened due to the co-polymerisation with vimentin when the F-actin crosslinking density is low compared to the network mesh size. Due to the changes in the network elasticity, the yield stress, and the strain stiffening, they suggest that this surprising emergent response comes from steric constraints on F-actin by vimentin (IF), promoting a lower degree of F-actin crosslinking in the final network.

The aim of this work is to develop a mechanical model capable to explain the observed rheological experiments performed by Jensen et al. \cite{jensen2014}. 
Interestingly, for the range of explored concentrations in the reported experiments, the vimentin network has a small role in the definition of the mechanical properties of the composite network, showing a high flexibility; but, nevertheless it plays a significant role setting physical crosslinks or steric constrains over the actin network. Then according to that, the main component of the structural mechanics will be given by the crosslinked F-actin network. Therefore, we develop a model defining an effective actin network which condenses the alteration of its structure on its main physical variables. In order to do so, we propose a mathematical model into the framework of non-linear continuum mechanics by using the semiflexible filament described by a worm-like chain following the Blundel-Terentjev formalism \cite{blundell2009}, and to homogenise the F-actin network we follow the 3-chain model as was implemented Meng et al \cite{meng2016, meng2017}. On the basis of this model, we introduce the dynamic effect of the crosslinks in order to capture the strengthening-weakening transition manifested by the network \cite{lopez2016, lopez2017, lopez2018}. Next, to capture the effects associated with the interaction F-actin/vimentin we propose an energy term associated with the interaction energy  by using the Landau model of phenomenological continuous phase transition where we define an interaction parameter that captures the effects of the alteration over the F-actin network due to the interaction \cite{nishimori2010, lopez2018}. Finally, we validate the model with experimental data coming from Jensen et al \cite{jensen2014}, and discuss the results and future works.

\section*{Methods}
In order to describe the theoretical constitutive model we first introduce the basic results associated with the framework non-linear continuum mechanics:
\subsubsection*{Basic results of the continuum mechanics} \label{sec_basic}
Let $\mathcal{B}_0$ be a continuum body defined as a set of points in a certain
assumed reference configuration. Denote by
$\{\boldsymbol{\chi}:\mathcal{B}_0\rightarrow\mathcal{R}^3\}$ the continuously
differentiable, one to one mapping (as well as its inverse
$\boldsymbol{\chi}^{-1}$) which puts into correspondence $\mathcal{B}_0$ with
some region $\mathcal{B}$, the deformed configuration, in the Euclidean space.
This one-to-one mapping $\boldsymbol{\chi}$ transforms a material point
$\textbf{X}\in\mathcal{B}_0$ to a position
$\textbf{x}=\boldsymbol{\chi}(\textbf{X})\in\mathcal{B}$ in the deformed
configuration.

The deformation gradient $\textbf{F}$ is defined as
\begin{equation}
    \textbf{F}:=\frac{\partial\boldsymbol{\chi}(\textbf{X})}{\partial\textbf{X}},
\end{equation}
with $J(\textbf{X})=\det(\textbf{F})>0$ the local volume ratio. It is sometimes useful to consider the multiplicative
split of $\mathbf{F}$
\begin{equation}\label{F_split}
    \textbf{F}=J^{1/3}\textbf{1}\bar{\textbf{F}},
\end{equation}
into dilatational and distortional (isochoric) parts, where $\textbf{1}$ is the
second-order identity tensor. Note that $\det(\bar{\textbf{F}})=1$. From this,
it is now possible to define the right and left Cauchy-Green deformation
tensors, $\textbf{C}$ and $\textbf{b}$ respectively, and their corresponding
isochoric counterparts $\bar{\textbf{C}}$ and $\bar{\textbf{b}}$
\begin{equation}\label{TensoresCyb}
\begin{array}{ll}
\textbf{C}=\textbf{F}^T\textbf{F}=J^{2/3}\bar{\textbf{C}}, &
\quad \bar{\textbf{C}}=\bar{\textbf{F}}^T\bar{\textbf{F}}, \\
\textbf{b}=\textbf{F}\textbf{F}^T=J^{2/3}\bar{\textbf{b}}, & \quad
\bar{\textbf{b}}=\bar{\textbf{F}}\bar{\textbf{F}}^T,
\end{array}
\end{equation}

For a hyperelastic material, the stress at a point
$\textbf{x}=\boldsymbol{\chi}(\textbf{X})$ is only a function of the
deformation gradient $\textbf{F}$ at that point. A change in stress obeys only
to a change in configuration. In addition, for isothermal and reversible
processes, there exists a scalar function, a strain energy function (SEF) $\Psi$, from which the
hyperelastic constitutive equations at each point $\textbf{X}$ can be derived. For materials with a particular symmetry group, the dependence of $\Psi$ on the deformation gradient is affected by the
symmetry group itself. Further, Spencer \cite{Spencer1980} showed that the irreducible integrity bases for the symmetric second-order tensors $\mathbf{C}$ and $\mathbf{a}_0\otimes\mathbf{a}_0$, correspond to four invariants:
\begin{equation}\label{invariantes}
\begin{array}{c}
  \begin{array}{ccccc}
  I_1=\mathrm{tr}\textbf{C}, &&
  I_2=\frac{1}{2}[(\mathrm{tr}\textbf{C})^2-
  \mathrm{tr}\textbf{C}^2], && I_3=
  \det\textbf{C}=1,
  \end{array}
\end{array}
\end{equation}
Invariants $I_1$, $I_2$, $I_3$ are standard invariants of the Cauchy-Green deformation
tensor, and are associated with the isotropic material behaviour. Invariant
$I_4$, arises from the anisotropy introduced by the remodelling. Next, it was proposed a representation of quasi-incompressible elasticity in which the SEF takes an
uncoupled form in which the dilatational and deviatoric parts are such that
\begin{equation}\label{SEF_anisotropic_uncoup}
\Psi(\textbf{X},\textbf{C},\textbf{a}_0)=U(J)+\bar{\Psi}(\textbf{X},
\bar{I}_1,\bar{I}_2,\bar{I}_4)
\end{equation}
where $\bar{I}_k, k=1,\ldots,4$, are the invariants of the isochoric
Cauchy-Green tensor $\bar{\textbf{C}}$ (note that $\bar{I}_3=1$). In the
developments in the next section, we use a SEF of the form given in Eq.~\ref{SEF_anisotropic_uncoup}.

For a hyperelastic material with a SEF, $\Psi$, defined above, the second Piola-Kirchhoff stress can be
written as:
\begin{equation}\label{piola_str1}
\mathbf{S}=2\frac{\partial\Psi}{\partial\mathbf{C}}=
pJ\textbf{C}^{-1}+2J^{-2/3}\mathrm{DEV}\left[\frac{\partial
\bar{\Psi}}{\partial \bar{\textbf{C}}}\right],
\end{equation}
where $p=U'(J)$, is the hydrostatic pressure, and $\mathrm{DEV}[\cdot]$ is the
deviatoric projection operator in the material description
\begin{equation}
\mathrm{DEV}[\cdot]\equiv[\cdot]-\frac{1}{3}([\cdot]:\bar{\textbf{C}})\bar{\textbf{C}}^{-1}.
\end{equation}
The Cauchy stress tensor is found by the weighted pushed forward of Eq.~\ref{piola_str1}
\begin{equation}\label{cauchy_str1}
\boldsymbol{\sigma}=J^{-1}\mathbf{FSF}^T=
p\textbf{1}+2J^{-1}\mathrm{dev}\left[\bar{\textbf{F}}\frac{\partial
\bar{\Psi}}{\partial \bar{\textbf{C}}}\bar{\textbf{F}}^T\right],
\end{equation}
where
\begin{equation}
\mathrm{dev}[\cdot]\equiv[\cdot]-\frac{1}{3}([\cdot]:\textbf{1})\textbf{1}.
\end{equation}

\subsection*{Free Energy}

In a first approximation we consider a Helmholtz free energy which accounts the strain energy functions associated with the elasticity of the crosslinked F-actin network and the intermediate filaments network made by vimentin, where its mechanical strain deformation is described by the Cauchy-Green tensor $\mathbf{C}$. Also, a last energy term associated with the interaction between networks; this term is proportional to the ratio between the concentration of actin and the concentration of vimentin $(c)$, this potential will allow us to define an interaction parameter $(\Gamma)$
\begin{equation}
\Psi(\mathbf{C},\Gamma,c)=\Psi_{IF}(\mathbf{\bar{C}},\Gamma)+\Psi_{actin}(\mathbf{\bar{C}},\Gamma)+\Psi_{inter}(c,\Gamma)+U(J), \label{FreeE}
\end{equation}
In the following we describe first the strain energy functions without considering the effects of the interaction defined by $\Gamma$; these effects will be defined further for clarity.

The first term in the free energy function refers to the intermediate filaments (IF). It has a long contour length and a low bending stiffness. They manifest a much higher flexibility in comparison with the F-actin network. In order to describe the soft mechanics of intermediate filaments we consider an isotropic Neo-Hookean strain energy function as: 
\begin{equation}
\Psi_{IF}(\bar{\mathbf{C}})=\frac{c_1}{2}\left(\bar{I}_1-3\right)
\end{equation}
where $c_1>0$ is a stiffness parameter.

The second term in the free energy function represents the strain energy function for the crosslinked actin network. This is modelled by means of a strain energy function (SEF) based on the wormlike chain model for semi-flexible filaments.  In order to do so, we propose a mathematical model into the framework of non-linear continuum mechanics by using the semiflexible filament described by a worm-like chain following the Blundel-Terentjev formalism \cite{blundell2009}. The two main physical parameters are the contour length of the filament, $L_c$, and the persistence length $l_p$, which represents a measure of the bundle stiffness and it compares the bending energy with the thermal energy, $l_p=EI/k_BT$. The chain is considered as semiflexible when $L_c \sim l_p$. Combining the effects of enthalpy arising from bending and entropy of conformational fluctuations, the closed form of the single chain free energy can be expressed as a function of its end-to-end factor, $x=\xi/L_c$:

\begin{equation}
\psi_{chain}=k_BT\pi^2\frac{l_p}{L_c}(1-x^2)+\frac{k_BT}{(1-x^2)}
\end{equation}

Next, we build the continuum elastic free energy of the network; in this sense several ways to perform the homogenisation have been proposed into the context of rubber and biopolymer based eight chains model, \cite{Arruda1993,Palmer2008,lopez2017,lopez2016} or by micro-sphere integration, \cite{ferreira2018}. Here, we choose to apply the three chain scheme, as was proposed by Meng et al, \cite{meng2016} because it allows the correct calculation of the normal stress.
The primitive cube for the homogenisation is constructed with lattice points representing the crosslink sites, and the edges are aligned along the principle directions of deformation tensor $\mathbf{C}$. Three chains are linked with their end-to-end vectors along the edges and the equilibrium mesh size $\xi$. On deformation, the lengths of the perpendicular edges over the lattice point become $\lambda_1 \xi, \lambda_2\xi$ and $\lambda_3\xi$ respectively. Then the free energy density of a semiflexible network can be expressed as:
\begin{equation}
\Psi_{3c}(\lambda_{i=1,2,3})=\frac{n}{3}\sum_{i=1,2,3}\psi_{chain}(\lambda_i\xi)
\end{equation}
\begin{equation}
\Psi_{3c}=\frac{nk_BT}{3}\left[(3-x^2I_1)+\frac{3-2I_1x^2+I_2x^4}{(1-I_1x^2+I_2x^4-I_3x^6)}\right]
\end{equation}

with $x=\xi/L_c$

If the stress tensor  is expressed as a function of the strain invariants for an incompressible material, where the $I_3=1$ can be expressed as:

\begin{equation}\label{s3s}
\mathbf{\sigma}=2\left[\left(\frac{\partial\Psi}{\partial \bar{I}_1}+\bar{I}_1\frac{\partial\Psi}{\partial \bar{I}_2}\right)\mathbf{\bar{C}}-\left(\bar{I}_1\frac{\partial\Psi}{\partial\bar{I}_1}+2\bar{I}_2\frac{\partial\Psi}{\partial\bar{I}_2}\right)\frac{\mathbf{I}}{3}-\frac{\partial\Psi}{\partial\bar{I}_2}\mathbf{\bar{C}.\bar{C}}\right]-p\mathbf{I}
\end{equation}

\begin{figure}[h]
	\includegraphics[width=13cm]{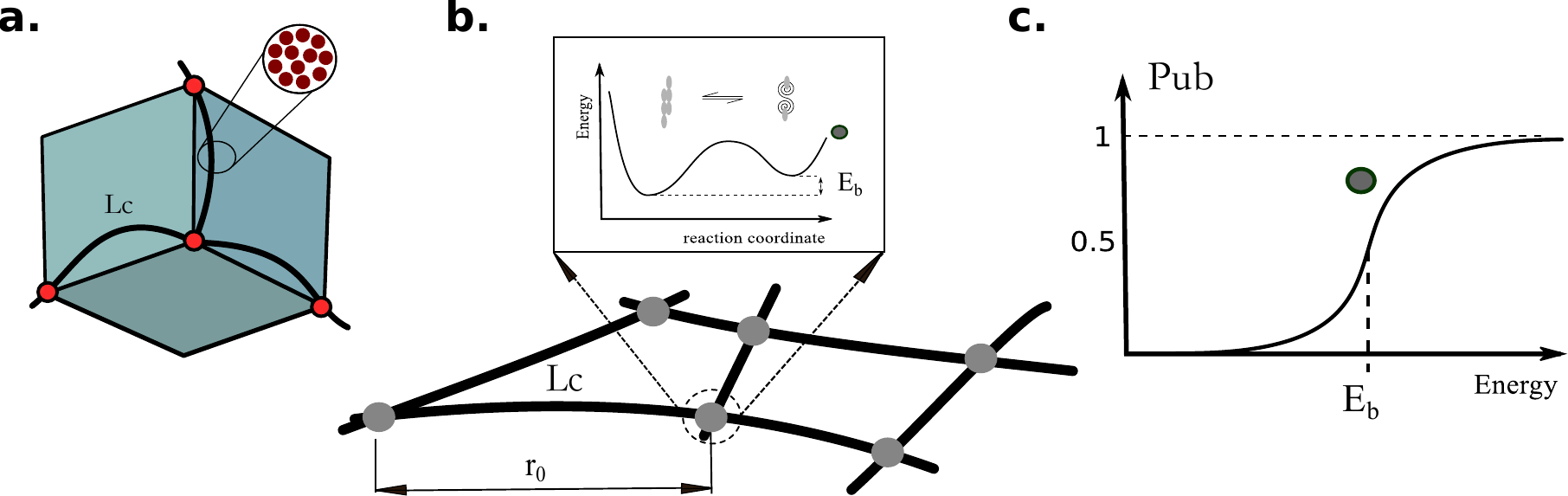}
	\centering
	\caption{(a) Three-chain homogenisation lattice. (b) filaments and crosslinks energy landscape. (c) unbinding probability $P_{ub}$. }
	\label{chainsfig}
\end{figure}

Formerly, we have described the constitutive model for a semiflexible network with rigid crosslinks (covalent bonds). In the following we address the necessary modifications into the model, in order to capture the network fluidisation due to transient crosslinks.

\paragraph*{Network fluidisation:}
The network is buildup by the interaction between the actin filaments and the crosslinks, and this defines the mechanical properties of the structure. If these interactions are stable (for the stress and the time scales of the experiments), it provides a strong gelation process and the network shows a solid-like behaviour under deformation. Nevertheless, many biological crosslink molecules have at least two properties that could cause network fluidisation: 
(i) force-induced unbinding and (ii) unfolding of the multiple internal protein domains that elongates the molecules. Previous studies have shown by computational simulation of the protein structure that it has flexible terminal regions which can twist and extend under mechanical stress without unbinding and lead to the lost of degrees of freedom \cite{Golji2009}. Then if the crosslinks are not completely stable, but they are associated with a reaction that can proceed in both directions, folding/unfolding, flexible/rigid states, binding/unbinding, we then speak of a weak gelation process,with the network showing a fluid-like behaviour and potentially without manifesting a complete unbinding.
As we mentioned previously the pre-strain into the structure produces an internal load for the bundle \cite{Lieleg2011}. Nevertheless, this also affects the crosslinks, where the level of pre-strain leaves them closer to the transition.
Then, to describe within the model the interaction between the crosslinks and the size of the mesh, i.e the contour length, $L_c$, we propose, in a similar manner as proposed Lopez-Menendez et al. \cite{lopez2016}, the following expression as:
\begin{equation}
L_{c}=L_{c}^{\min}+\delta L_c^{cl}P_{ub},
\end{equation}
where $P_{ub}$ defines the unfolding probability encompassing the states of unfolding or flexible crosslink, $L_{c}^{\min}$ represents the contour length when $P_{ub}=0$ (folded crosslink), and $\delta L_c^{cl}$
represents the average increment of the contour length when the unbinding probability is one.

In this sort of networks the chemical crosslinks are not covalent bonds with high adhesion energy, in fact their adhesion energy is in the order of tens of $k_BT$, an having a transient dynamics \cite{ferrer2008}. In general terms, this kind of gels with chemical crosslinks (proteins as $\alpha$-actinin) behaves as a physical gels \cite{de1979}. This sort of interactions can be modelled as  a reversible two-state equilibrium process \cite{Brown2009,lieleg2010,purohit2011}. Moreover, taking into account that the shear velocity is much slower than the internal crosslinks dynamics we can consider the interaction at steady state. Then the process can be described as:
\begin{equation}
\frac { P_{ ub} }{ P_{ b} } =\exp -\frac {( E_b - w_{ ext} )}{ k_{ B }T },  \label{eq_7}
\end{equation}
where $P_b$ the binding probability encompassing the states folding or rigid crosslink. Since only these two states are possible, then $P_{ub} + P_b = 1$. The two-state model has the folded state as the preferred low free energy equilibrium state at zero force and the unfolded state as the high free energy equilibrium state at zero force. $E_b$  represents the difference in the free-energy between these states.  $w_{ext}$ represents the external mechanical work that induces the deformation of the crosslink.

As we are developing a mesoscale model, in the following we write an expression for the unbinding probability considering the shear strain as the main driving force, by using scaling arguments \cite{Bell1978,de1979}. Then, in order to do so  we can re-write as  $w_{ext}=f.a$, where $a$ is a length scale in the order of the monomer size. The force $f$ can be expressed as $f\sim G\gamma \xi^2$ in which $\gamma$ is the shear strain, $G$ is the shear modulus which can be estimated as $G\sim\frac{l_p k_BT}{L_c\xi^3}$, and $\xi$ the network mesh size. Also, taking into account that the unbinding transition due to the bundle strain happens in the semiflexible regime when $\xi \sim L_c$. Therefore, reorganising the terms we arrive to an expression for the $P_{ub}$ as a function of the shear strain as: 
\begin{equation}\label{eq_pub}
P_{ub}=\frac{1}{1+\exp\left[\kappa\left(\gamma_{0}-\gamma\right)\right]}, \quad
\gamma_0\sim\left(\frac{E_b \xi^2}{k_BT l_p a}\right)
\end{equation}
where the parameter $\kappa\sim\frac{l_p a}{\xi^2}$ is proportional to the sharpness of the transition between states and $\gamma_0$ is the characteristic strain which is proportional to the adhesion energy; it defines the point at which the probability of unbinding is 0.5. If $\gamma_{0} << \gamma$, the network is easy to be remodelled showing a fluid-like behaviour. On the contrary, if $\gamma_{0} >> \gamma$, the crosslinks are stable and the probability of transition is low, consequently the network behaves as a solid-like structure. Moreover, we can clearly identify that the characteristic strain $\gamma_0$, scales proportionally with the adhesion energy $E_b$, with the mesh size $\xi$ and increases when the bundle stiffness $l_p$, becomes smaller.
\begin{figure}[h]
	\includegraphics[width=14cm]{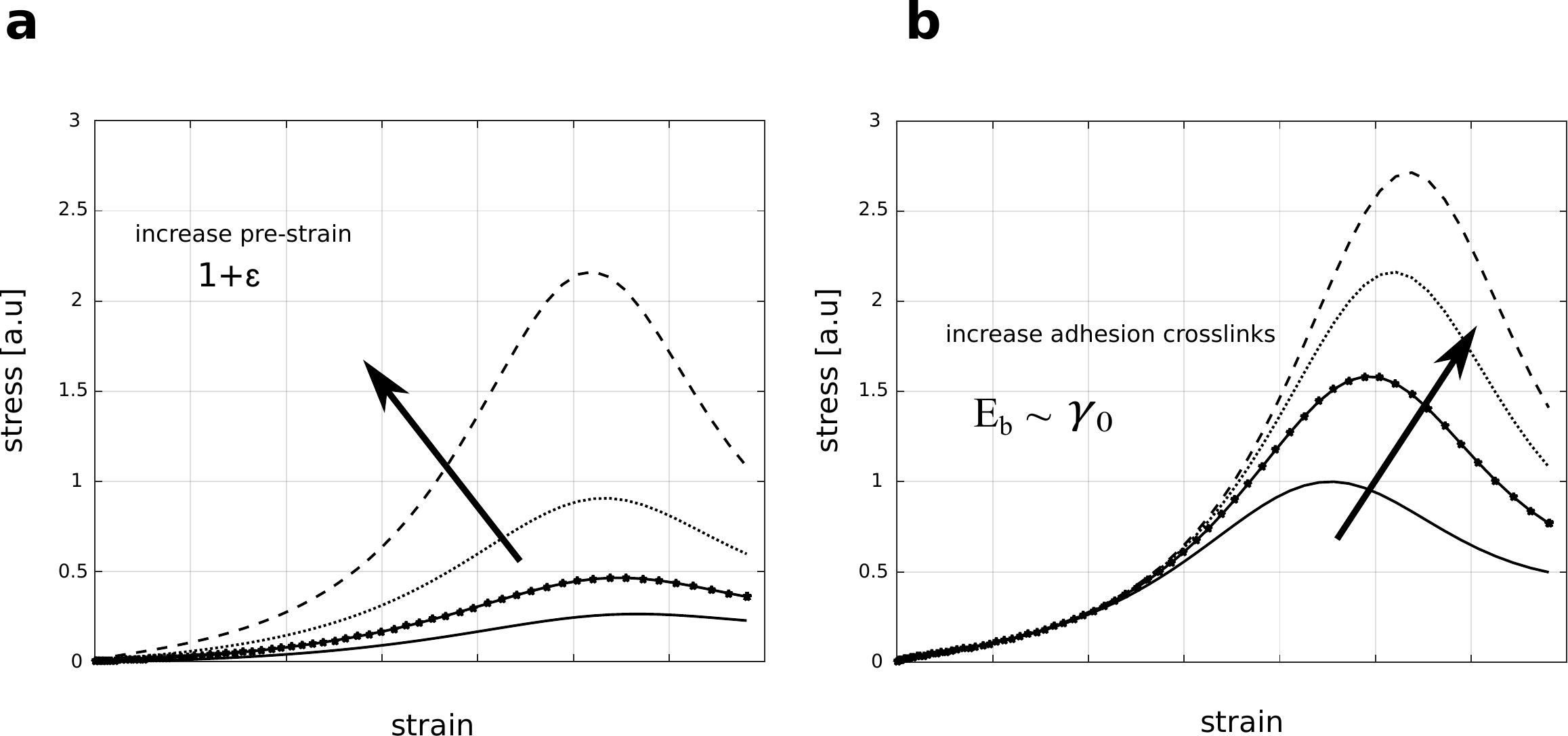}
	\centering
	\caption{(a) Describes the network response for different levels of pre-strain. It shows the increment in the slope for low values of network deformation. (b) The figure details the network response for increasing values of $\gamma_0$ showing an extension of the solid-like regime.}
	\label{stress_ways}
\end{figure}

In order to express qualitatively the behaviour of the coupled set of equations under alterations in the pre-strain and the adhesion energy of the crosslinks, we evaluate them in the regime of semi-flexible response i.e $L_c \propto l_p$. 
Figure \ref{stress_ways}.a describes the effect of an increment on the pre-strain (1+$\epsilon$) on network response, with the remaining parameters keeping constant. As can be observed, as the pre-strain increases, the network stiffness increases and is able to reach a higher level of stress (higher yield point). However, the yielding point (fluidisation of the network) occurs earlier reducing the solid-like regime of the network. Figure \ref{stress_ways}.b shows the response of the network for different values of $\gamma_0$. Contrary to the pre-strain, as $\gamma_0$ increases the initial stiffness of the network remains unaltered while the yielding stress and strain increase, extending the solid-like regime. This implies that as $\gamma_0$ increases the crosslinks become more stable. 

\subsection*{Interaction between actin and vimentin}

In order to consider the interaction between networks we expect that for very low concentrations the changes over the mechanical response of the actin network is almost negligible, but once a certain value is overpassed the effects associated with the interaction are most relevant until some asymptotic value. This effect can be interpreted by using arguments from phase transition, according to that we propose to use an interaction energy $\Psi_{int}$ by means the Landau functional that couple the effects with the networks \cite{nishimori2010, lopez2018}. This energy is written in terms of an interaction parameter defined $\Gamma=\Gamma(c)$ where $c$ represents the ratio between the concentrations of actin and vimentin.  
As we are interested to know when the effect of the vimentin (IF) becomes relevant on the mechanical response we focus on the critical phenomena, when the concentration of $c$ is near the critical point and the interaction parameter $\Gamma$ assumes a very small value. 
This allows us to expand the free energy in even powers of $\Gamma$ and retain only the lowest order terms. 
 
Then we re-write the Helmholtz free energy as follows:
\begin{equation}
\Psi( \bar{\mathbf{C}}, \Gamma)=\frac{\alpha}{2}\Gamma^2 + \frac{\beta}{4}\Gamma^4 + \Psi_{actin}(\bar{ \mathbf{C}}, \Gamma) + \Psi_{IF}(\bar{\mathbf{C}})
\end{equation}
Where the first two terms define the Landau energy associated with the interaction parameter; the third term interprets the strain energy for the network without any coupling as a function of the isochoric Cauchy strain tensor and the interaction parameter.  Since the equilibrium position (minimum) of $\Psi(\Gamma,c)$ changes at $\alpha=0$, we identify $\alpha=0$ with the critical point $c=c_{cr}$. It allows us to choose $m\hat{c}$ as $\alpha$, where $m$ is a positive constant and $\hat{c}=(c-c_{cr})/c_{cr}$ is the deviation of the concentration ratio from the critical point normalised by $c_{cr}$ which we define as a reduced concentration ratio. Then, the simplest election is $\alpha=k\hat{c}$, for which $\alpha>0$ above the critical point and $\alpha<0$ below. The dependence of $\beta$ with $\hat{c}$ does not affect qualitative the behaviour of the free energy in the vicinity of the critical point and therefore we take $\beta$ as a constant. Then minimising the free energy to obtain the equilibrium condition with respect to the interaction parameter $\Gamma$ yields:
\begin{equation}
\frac{\partial \Psi}{\partial \Gamma}\approx2\alpha\Gamma+4\beta\Gamma^3=0.
\end{equation}
Thus, the equilibrium value of remodelling, $\Gamma$ is
\begin{equation}
\Gamma\approx\left( \frac{-\alpha}{2\beta}\right)^{1/2} =\left\{ 
\begin{array}{rcl}
  \left[\frac{m(c-c_{cr})}{2\beta c_{cr}} \right]^{1/2}  &\forall &  c>c_{cr} \\
  0 &\forall \quad & c<c_{cr}		
\end{array}
\right.
\end{equation}
The interaction parameter is canceled when the concentration ratio $c \approx c_{cr}$, above the critical value scale as $\Gamma \sim (c-c_{cr})^{\frac{1}{2}}$. For values of $c$ below the critical the level of interaction is zero. Then, once the interaction parameter has been defined, we will describe in the following, the internal variables that encode the interplay between the two networks and how they are driven by the interaction parameter $\Gamma(c)$. We consider the following hypothesis:
\subsubsection*{Interaction induce strengthening:}

\begin{figure*}[h]
	\centering
	\includegraphics[width=13cm]{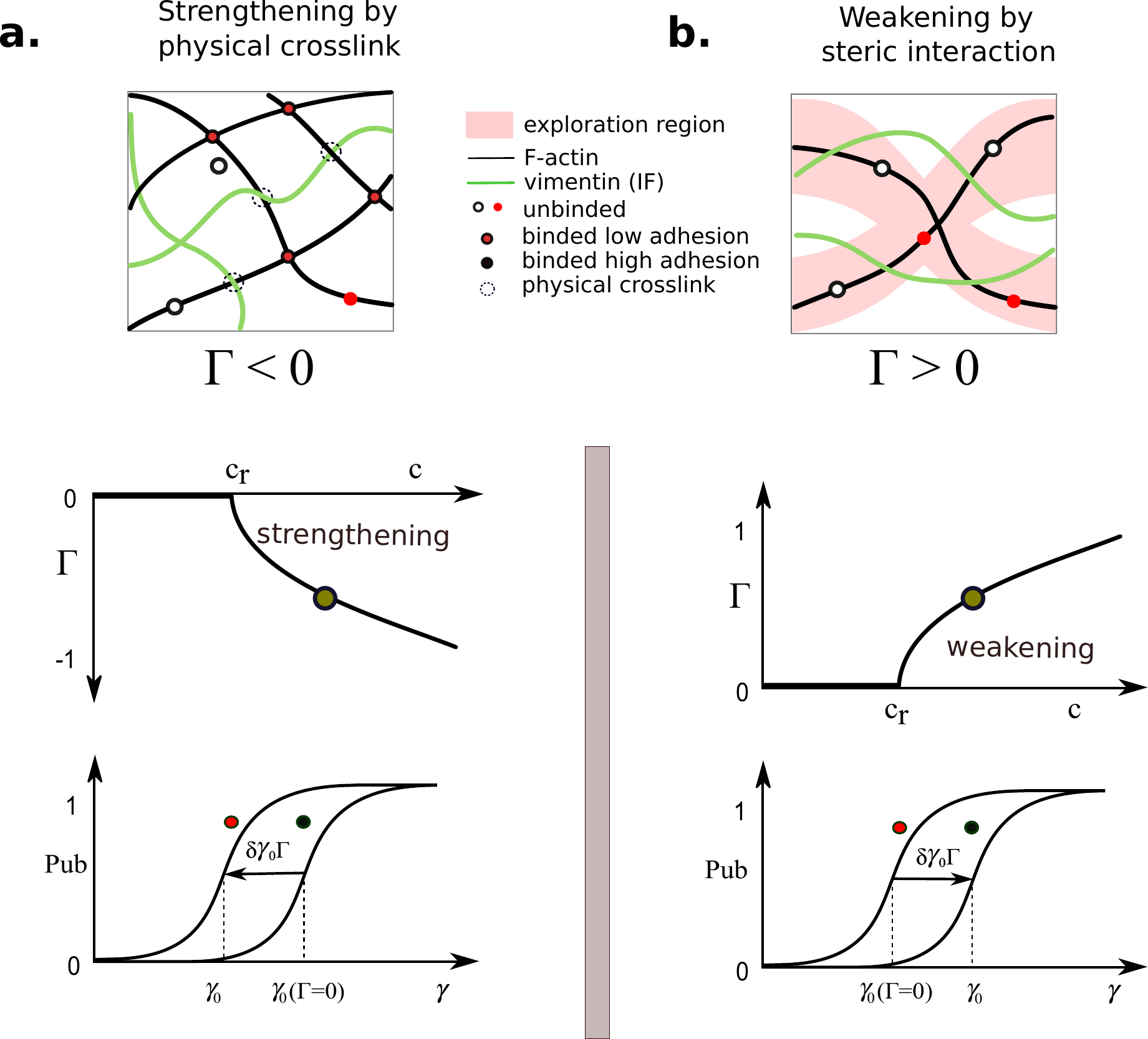}	
	\caption{(a) Strengthening promoted by the formation of physical crosslinks. (b) Weakening promoted by the steric interaction that disturb the formation of chemical crosslinks, which increase the contour length.}
	\label{quant_resp}
\end{figure*}

 \textbf{i.} The increment in the concentration of IF promotes an increment of the physical crosslinks over the F-actin bundles reducing the contour length and reducing the degree of fluctuations of the actin. As the $L_c$ is reduced the ratio $r/L_c$ tends to one and the composite network manifests a rise in the stress. Nevertheless, as the IF filaments are very flexible, the increment in the density of physical crosslinks due to the interaction with F-actin does not produce a relevant change over the stress sustained by the network. Therefore, in order to simplify the model we neglect the effect of the physical crosslinks over the IF and only focus on the role of physical crosslinks over the F-actin, as can be observed in the figure~\ref{quant_resp}a (top). Then finally the effective contour length $L_c$ due to the alterations of vimentin.
\begin{equation}
L_{c}(\Gamma)=L_{c}^0-\delta L_{c}^{\Gamma}\Gamma+\delta L_{c}^{cl}P_{ub},
\end{equation}
where $\delta L_{c}^{\Gamma}\Gamma$ represents the reduction of the contour length due to vimentin interaction,
as was outlined previously $L_{c}^0$ represents the contour length of the for the mesh without vimentin. The second term $\delta L_{c}^{\Gamma}\Gamma$ represents the effective reduction in the length associated with the formation of the physical crosslinks promoted by the vimentin (figure \ref{quant_resp}a.). 

\textbf{ii.} The effective network is a representation of a network buildup by two kinds of transient crosslinks. On the one hand the chemical interactions given by the crosslinks of neutravidin; on the other hand the physical crosslinks due to the interaction between F-actin with vimentin. Therefore, we expect that the effective $\gamma_0$ will be smaller because it represents a lower effective adhesion energy, which is proportional to the mixture of physical and chemical crosslinks. The adhesion energy promoted by the physical crosslinks and given by friction among filaments (without strong entanglements), is lower, in comparison with the chemical crosslinks $E_b$ \cite{de1979}. Moreover, another effect that promotes the reduction of the yielding strain is associated with the fact that the rise of the internal stress, associated with the physical crosslinks is propagated towards the chemical crosslinks lowering the characteristic strain $\gamma_0$, as shown in the figure~\ref{quant_resp}a. (bottom part), where the red and black dots illustrate the effect of the pre-stress over the $P_{ub}$ of the chemical crosslinks \cite{Lieleg2009,Lieleg2011}. 

Hence from the perspective of the proposed model the changes induced by the increment of the density of IF in the F-actin network are encoded as decreases in $L_c,$ and in $\gamma_0$. Based on that we write them as a combination of the parameters associated in a network without IF plus a perturbation associated with the interaction parameter $\Gamma$, as follows: 
\begin{equation}
\gamma_0(\Gamma)=\bar{\gamma_0}-\delta\gamma_0\Gamma,
\end{equation}

\subsubsection*{Interaction induce weakening:}

 \textbf{i.}  Surprisingly, at high actin concentrations, the additional polymer results in an unexpectedly weaker composite network with a lower elasticity and yield stress. The crossover between the strengthening and weakening regimes observed in the composite network, occurs when the estimated F-actin network mesh size is comparable to the distance between F-actin crosslinking sites. When the actin concentration increases and the concentrations of crosslinks and vimentin are the same as in the strengthening experiments, the ratio between $\chi=\frac{[crosslinks]}{[actin]}$ is lowered and the resultant mesh size increases as well as the level of thermal fluctuations.  Consequently, the probability of bond formation is lower. In addition to that when the networks are co-polymerised with these ratio crosslinks / actin and with the range of concentrations of vimentin. The interaction disturbs the crosslinking process providing an additional steric constraint imposed by vimentin IF which results in a loss of F-actin crosslinking. Based on that we write them as combination of the parameters associated in a network without IF plus a perturbation associated with the interaction parameter $\Gamma$, as follows: 
 
 \begin{equation}
L_{c}(\Gamma)=L_{c}^0+\delta L_{c}^{\Gamma}\Gamma+\delta L_{c}^{cl}P_{ub},
\end{equation}
where as was previously described  $L_{c}^0$ represents the contour length of the mesh without vimentin. The second term $\delta L_{c}^{\Gamma}\Gamma$ represents the effective increase in the length associated with the steric interaction promoted by the vimentin. 
 
\textbf{ii.} The rise of the internal stress is propagated towards the chemical crosslinks lowering the characteristic strain $\gamma_0$, as we shown in the figure~\ref{quant_resp} where the red and black dots describe the effect of the pre-stress over the $P_{ub}$ of the chemical crosslinks \cite{Lieleg2009,Lieleg2011}. Hence, from the perspective of the proposed model, the changes induced by the increment of the density of IF in the F-actin network are encoded as a decrease in $\gamma_0$. 
\begin{equation}
\gamma_0(\Gamma)=\bar{\gamma_0}+\delta\gamma_0\Gamma,
\end{equation}

\section*{Results}
The proposed theory is used to depict the experiments conducted by Jensen et al.\cite{jensen2014} on copolymerised F-actin/vimentin network. We evaluate the proposed model for the set of parameters identified by means of nonlinear least-square fit with experiments of monotonic shear tests, in a regime of large deformation, as is reported in~\cite{jensen2014}. Subsequently, solving the following coupled set of equations we can obtain the stress-strain relation for the different analysed networks:
\begin{equation}\label{eqLam}
\gamma_0(c)=\bar{\gamma_0} \pm \delta \gamma_0\left[\frac{ m(c_{cr}-c)}{2\beta c_{cr}} \right]^{1/2} 
\end{equation}
where $\pm$, as was explained above, will depend on the actin concentration. Next, the contour length is:
\begin{equation}\label{eqLc}
L_{c}(c,\gamma)=L_{c}^{0}\pm\delta L_{c}^{\Gamma}\left[\frac{m(c_{cr}-c)}{2\beta c_{cr}} \right]^{1/2} + \frac{\delta L_c^{cl}}{1+\exp\left[\kappa\left(\gamma_{0}-\gamma\right)\right]}. 
\end{equation}
where the updated mesh for the reference configuration becomes
\begin{equation}
x(L_c)=(1+\epsilon)\left(1-\frac{2L_{c}(c,\gamma)}{l_p \pi^{\frac{3}{2}}}\right)^{\frac{1}{2}},     
\end{equation}
Finally rewriting the eq.\ref{cauchy_str1} and the eq.\ref{s3s} considering that the incompressibility is satisfied automatically and the remaining of the invariants are: $I_1=I_2=3+\gamma^2$ we obtain the new expression for the shear stress as:
\begin{equation}
\sigma_{xz}(\gamma)=c_1\gamma+\frac{2}{3}nk_BT\gamma x^2 \left[
\frac{(1-x^2)}
{c \pi \left[1-(2+\gamma^2)x^2+x^4\right]^2}-c\pi^2
\right]
\end{equation}

The parameters of the model can be divided in two types: (i) Rigid-wormlike chain parameters $L_{c}^{0}$, $l_{p}$, $\delta L_{c}^{\Gamma}$, $\delta L_{c}^{cl}$ and $\epsilon$ which are of the order of magnitude of the values used to describe in experiments of in-vitro F-actin networks and to keep on the regime of semi-flexible entropic elasticity \cite{Palmer2008, meng2016,lopez2016}.  (ii) The parameters associated with the remodeling dynamics of the crosslinks $\kappa$ and $\gamma_{0}$, and the parameters that describe the interaction parameter $\Gamma(c)$. These parameters encode the transitions to induce the fluidisation of the network and represent an indirect measure of the adhesion force of crosslinks.These values were identified in order to fit the experimental data.

\begin{figure}[h]
	\includegraphics[width=14cm]{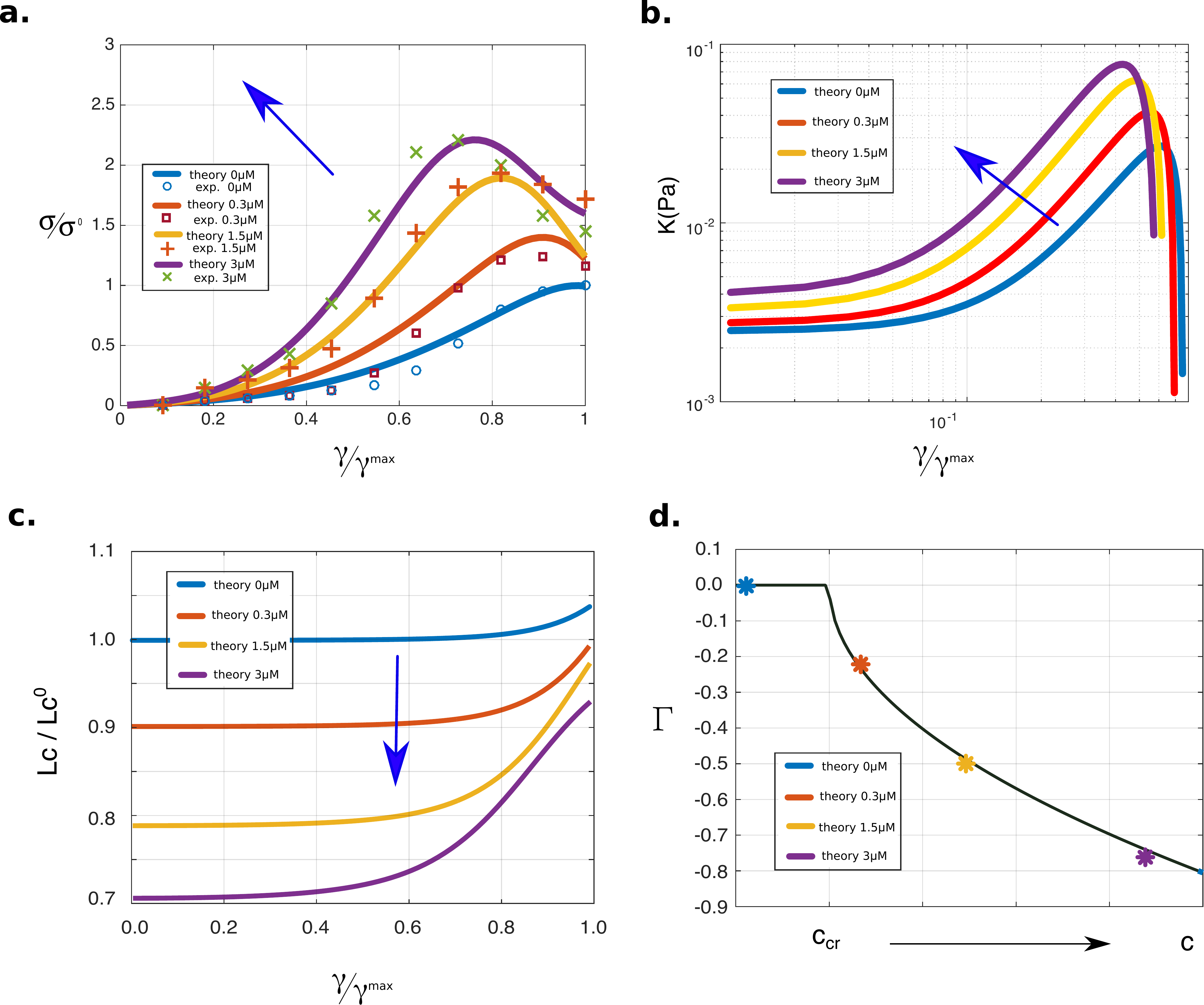}
	\centering
	\caption{Strengthening effect. (a) stress-strain plots under shear strain regime, showing the non-linear inelastic effects. It can be observed a good agreement between the model predictions and the experimental measurements. (b) Effect of initial strengthening is illustrated by $K=\frac{d\sigma}{d\gamma}$, the blue arrow point the direction of the strengthening increase. (c) $\frac{L_c}{L_c^0}$ for different concentrations of vimentin and shear strain; the figure condenses the effects associated with the interaction parameter $\Gamma$ and crosslinks fluidisation. (d) interaction parameter $\Gamma(c)$ as a function of the concentrations ratio c=[vimentin]/[F-actin].}
	\label{stiff}
\end{figure}

\subsubsection{strengthening phase:}

The strengthening effect is a consequence of the formation of physical crosslinks. In this case, the concentration of F-actin keeps constant at $6\mu M$ and the vimentin encompasses in the range: $0\mu M, 0.3\mu M, 1.5 \mu M, 3 \mu M$. Then, in the figure \ref{stiff}a. we plot the model predictions and the experimental measurements from Jensen et al. for the stress-strain curve of the composite actin-vimentin network under the application of a simple shear \cite{jensen2014}. We notice that the model is capable of capture the general trend of the experimental results, associated with the strengthening as well as the increment of the $\sigma_{max}$ and the reduction of the $\gamma_c$ when the concentration of vimentin (intermediate filaments) increases. Moreover, to better illustrate the increment of the linear modulus due to the presence of vimentin, we plot in the figure \ref{stiff}b. the modulus $K=\frac{d\sigma}{d\gamma}$. It can clearly be observed that the value of $G_0\approx K_{\gamma=0}$ rises with the concentration of vimentin.

Furthermore, to better characterise the alterations over the mesh size on the effective network we illustrate figure \ref{stiff}c., showing the changes over the contour length $L_c$, due to the strain $\gamma$ and the interaction parameter $\Gamma(c)$. As described above with the eq.$\ref{eqLc}$, the $L_c$ is reduced by the term $\delta L_{c}^{\Gamma} \Gamma$, this express the increment of the density of the physical crosslinks. The reached reduction can be in the order of $30\%$ with respect to the contour length $L_{c}^0$, without vimentin. In addition to that, the second term into eq.$\ref{eqLc}$ describes the the increment in the contour length due to the rise of $\gamma$, which finally enhances the unbinding probability, $P_{ub}(\gamma)$. As can be also observed in the figure, if the concentration of vimentin increases, the effects of crosslinks fluidisation becomes more relevant. This is due to the negative coupling between $\Gamma$ and $\gamma_0$ following the eq.$\ref{eqLam}$. Finally, with the figure \ref{stiff}.d we can observe the functional form of the the interaction parameter $\Gamma(c)$ as a function of the ratio of concentrations $c$ where the points express the associated values of the ratio $c=\frac{[vimentin]}{[F-actin]}$ and $\Gamma$ that allow finally the described stress-strain curves.

\begin{figure}[h]
	\includegraphics[width=14cm]{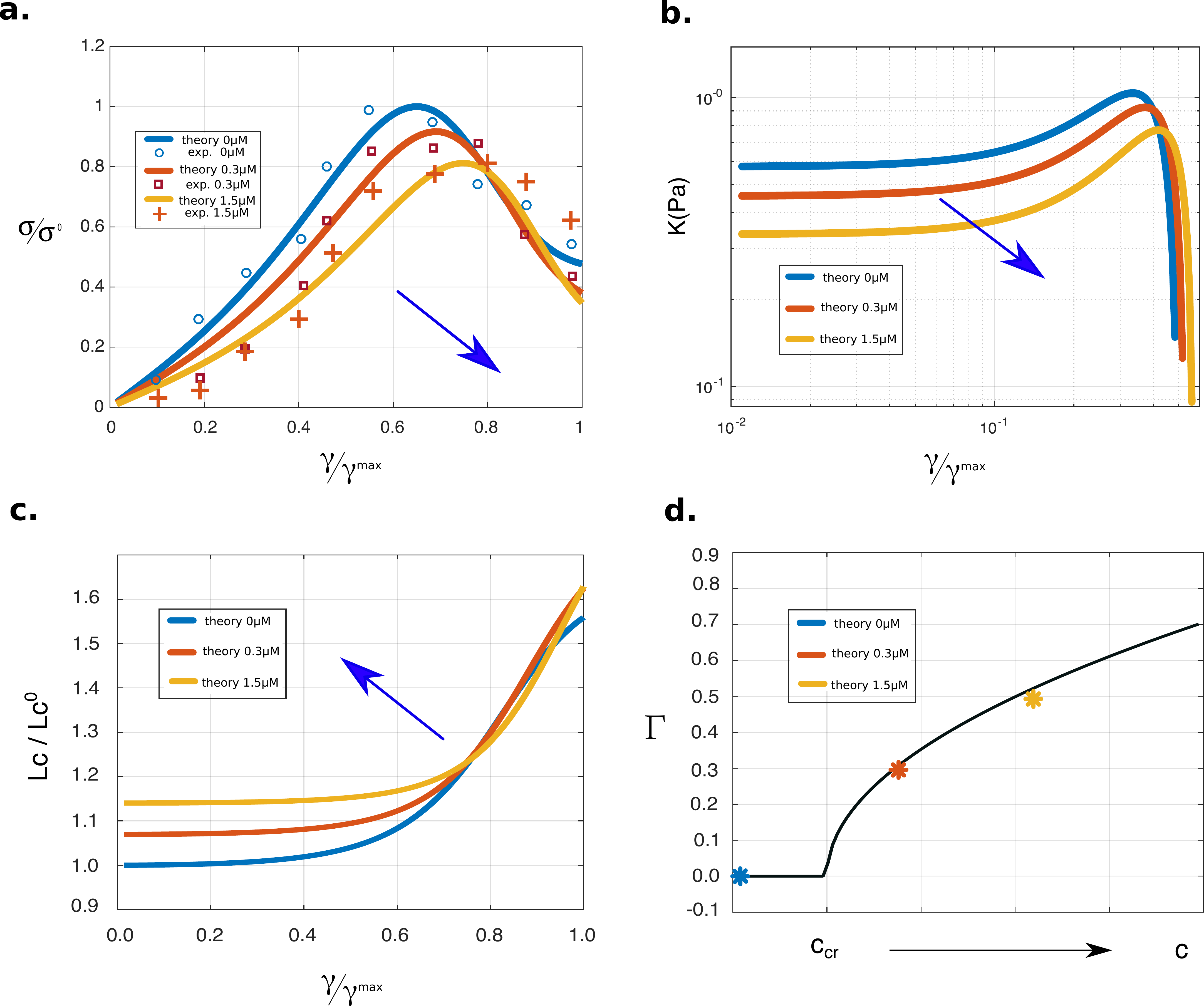}
	\centering
	\caption{Weakening effect into composite networks F-actin/vimentin where the copolymerisation promotes steric interactions that reduce the formation of crosslinks. In this case the concentration of actin keeps constant and the vimentin rises in the range $(0\mu M, 0.3\mu M, 1.5 \mu M, 3 \mu M)$. (a) stress-strain curves under shear strain loading regime showing the non-linear elastic effects. It can be observed a good agreement between the model predictions and the experimental measurements. (b) The effect of the initial weakening is illustrated by $K=\frac{d\sigma}{d\gamma}$, the blue arrow points the direction of the stiffening decrease. (c) Alterations in the contour length due to the steric interaction that reduce the chance of crosslinks formation, increasing the mesh size (blue arrow). (d) Interaction parameter $\Gamma(c)$ as a function of the concentrations ratio.}
	\label{soft}
\end{figure}

The scaled material parameters for the simulation are: $\frac{\delta L_c^{\Gamma}}{L_c^0}=0.4$; $\frac{\delta L_c^{cl}}{L_c^0+\delta L_c^{\Gamma}}=0.5$; $\frac{l_p}{L_c^0}=0.8$; $\frac{\bar{\gamma_0}}{\gamma^{max}}=0.9$; $\frac{\delta\gamma_0}{\gamma^{max}}=0.4$; $\frac{c_1}{\sigma^0}=0.2$; $\kappa=30$; $\frac{m}{2\beta}=0.5$; $\epsilon=0.03$.

\subsubsection{weakening phase:} In the following we describe the results provided by the model to capture the experimentally reported emergent softening phase into the composite networks F-actin/ vimentin which promotes a steric interaction that blocks the formation of crosslinks. The studied concentration of F-actin kept constant at $18\mu M$ and the vimentin encompasses in the range: $0\mu M, 0.3\mu M, 1.5 \mu M, 3 \mu M$. In order to show the results of the weakening, in the figure \ref{soft}a. we plot the model predictions and the experimental measurements from Jensen et al. for the stress-strain curve of the composite actin-vimentin network. The model is capable of capture the general trend of the experimental results, associated with the weakening as well as the reduction of the $\sigma_{max}$ and the increment of the $\gamma_c$ when the concentration of vimentin (intermediate filaments) increases. Moreover, to better illustrates the increment of the linear modulus due to the presence of vimentin we plot in the figure \ref{soft}b. the modulus $K=\frac{d\sigma}{d\gamma}$. It can clearly be observed that the value of $G_0\approx K_{\gamma=0}$ rises with the concentration of vimentin. The alterations over the mesh size on the effective network we illustrate in figure \ref{soft}c., showing the changes, due to the strain $\gamma$ and the interaction parameter $\Gamma(c)$, over the contour length $L_c$ respect the contour length for a mesh without vimentin $L_c^0$. 
 The eq.$\ref{eqLc}$, the $L_c$ is increased by the term $\delta L_{c}^{\Gamma} \Gamma$, which describes the steric interaction, where the increment can be in the order of $10\%$ with respect to the contour length $L_{c}^0$, without vimentin. Next, the second term into eq.$\ref{eqLc}$ depicts the extension of the contour length due to the increment in $\gamma$ which enhances the unbinding probability, $P_{ub}(\gamma)$, where $\gamma_0$ denotes the crosslinks fluidisation transition.  As can also be observed \ref{soft}.c if the concentration of vimentin increases the effects of crosslinks stabilisation becomes more relevant. This effect is due to the positive coupling between $\Gamma$ and $\gamma_0$ following the eq.$\ref{eqLam}$. Lastly, the figure \ref{soft}d. depicts the positive functional form of the interaction parameter $\Gamma(c)$ as a function of the ratio of concentrations $c$. The points express the associated values of the ratio $c=\frac{[vimentin]}{[F-actin]}$ and $\Gamma$ the described stress-strain curves.

\begin{figure}[h]
	\includegraphics[width=10cm]{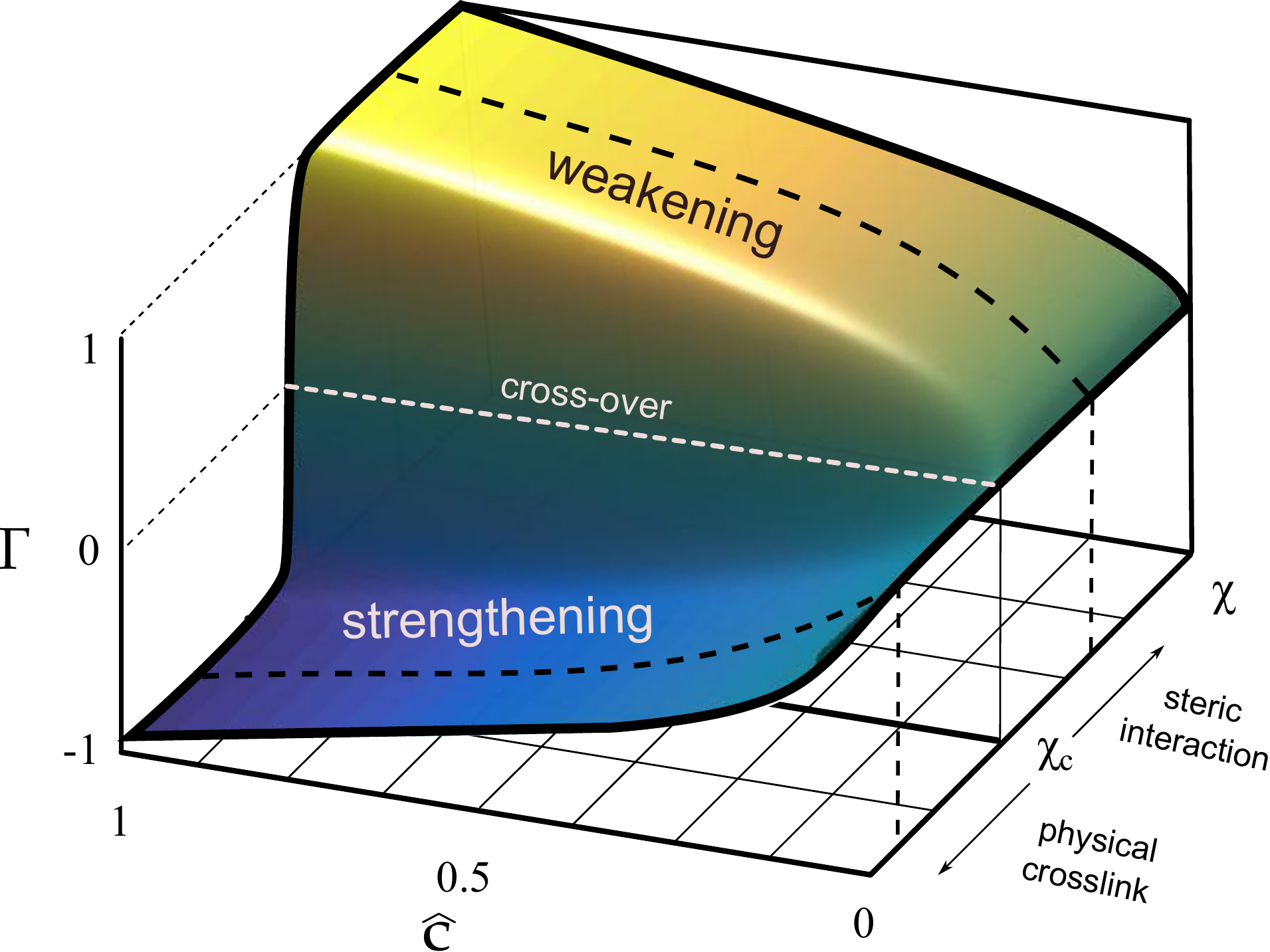}
	\centering
	\caption{Proposed phase diagram to describe the effect of strengthening and weakening}
	\label{phase_d}
\end{figure}

The scaled material parameters for the simulation are: $\frac{\delta L_c^{\Gamma}}{L_c^0}=0.4$; $\frac{\delta L_c^{cl}}{L_c^0+\delta L_c^{\Gamma}}=0.6$; $\frac{l_p}{L_c^0}=1.2$; $\frac{\bar{\gamma_0}}{\gamma^{max}}=0.68$; $\frac{\delta\gamma_0}{\gamma^{max}}=0.4$; $\frac{c_1}{\sigma^0}=0.2$; $\kappa=60$; $\frac{m}{2\beta}=0.45$; $\epsilon=0.02$.

\section*{Discussion and Conclusions}
In summary we provide in this work a first study of a constitutive model for composite networks of crosslinked F-actin/vimentin. It was motivated by the fact that some previous rheological measurements on composite biopolymer networks, such as F-actin/microtubules, showed that the composite networks always induce strain strenghening in comparison with the single F-actin network.  Nevertheless, the experiments of Jensen et al. demonstrated that the composite semiflexible networks of F-actin/vimentin can drive either the mechanical strengthening or weakening, during the co-polymerisation of the two semiflexible species. 

The model has successfully reproduced the experimental observations. More importantly, it can be readily implemented into a field theory and used to calculate the behaviour of a composite networks of actin/vimentin under complex loading conditions. Our theory was developed into the framework of nonlinear continuum mechanics, in which we define a free energy functional considering the role of the entropic-elastic for semiflexible networks with weak crosslinks and also an energetic term to describe the interaction parameter, which allows the coupling between the two networks.  Surprisingly, our phenomenological approach provides a very simple and useful constitutive model, which can capture the two described mechanisms of strengthening and softening just as a change in the sign of the interaction parameter $\Gamma(c)$.

This effect leads us to think that the formation of the cytoskeleton scaffolding elements can drive to a broad phase diagram for the cellular mechanical properties. In this sense, the figure \ref{phase_d} condenses our interpretation of the process. We consider that the effects of strengthening and weakening can be considered as the action of two ratios of concentrations (which could also be described as chemical potentials), one is defined as $c=\frac{[actin]}{[vimentin]}$, and the other as $\chi=\frac{[crosslinks]}{[actin]}$. The first one, $c$ will define the intensity of the interaction, where we find a phase transition in which above a certain critical value the coupling between networks becomes more dominant. The second concentration ratio controls the sort of interaction. For $\chi=\chi_c$ exists a crossover between the two regimes: 
On the one hand, below the crossover $\chi<\chi_c$the strengthening, where the interaction creates physical crosslinks, in which the effective network has a reduced contour length. Moreover the effective adhesion energy ($\propto \gamma_0$), is a weighting between the chemical crosslinks and the new physical crosslinks, then seems plausible to expect that the yielding strain decreases. In addition to that, the rise of the physical crosslinks makes higher the network pre-strain, and consequently the mechanical stress over the chemical crosslinks which reduce the $\gamma_0$.  On the other hand, above the crossover $\chi>\chi_c$ the formation of the transient chemical crosslinks becomes scare, which increases the mesh size. On that condition, the only way to raise the chance of crosslinks formation is by the rise of the level of fluctuations. Nevertheless, the co-polymerisation with vimentin reduces the level of the internal fluctuations and consequently the effective mesh size becomes smaller.  Furthermore, as the mesh size becomes higher, the level of pre-strain over the crosslinks becomes smaller and the interactions do not reduce the adhesion energy. Therefore it explains why the yielding strain becomes higher.

Taking all the observations as a whole, we propose a phase diagram where the coupling between $\chi$ and $\Gamma$ could have a functional form as $\sim \tanh(\chi-\chi_c)$ (see figure \ref{phase_d}). Thus, it allows the change of the sign in the interaction parameter depending on whether the value is above or below the crossover $\chi_c$. Future experiments will provide better arguments to validate the speculative relation. Essentially we propose the use of an effective crosslinked F-actin network, which incorporates all the associated actin/vimentin interactions that drives the microstructural remodelling effects via the alterations of the contour length $L_c$, and the characteristic stretch $\gamma_0$. Taking a broad perspective, several models have described the effects of composite materials where one is considered the most relevant and the other one is considered as the surrounding matrix. Generally, in all these models the coupling between the two components enhances the strengthening, but never the weakening \cite{JFRodriguez}. This approach has also been used in different studies of composite materials which address the mechanical interactions between filaments of fibres with the surrounding matrix. In this sense the formalism developed by Winkler accounts for coupling with an elastic foundation that resists a lateral displacement of a slender structure. This approach was used to describe the alterations in the mechanical response of the microtubules due to the effects produced by the surrounding actin network \cite{lee2018,brangwynne2006}. This kind of interaction is studied by the modification of the buckling modes. Lee and Terentjev describe this interplay by means a partition function which considers a Hamiltonian associated with the bending energy of the microtubule, plus the Winkler interaction energy due to the physical constrains, introduced by the actin filaments \cite{lee2018}. Future works exploring the interaction between intermediate filaments and F-actin could gain novel insights by using a similar techniques.

In a similar manner, the definition of an effective actin network that considers the interplay with vimentin can be seen as Winkler-like model. Our methodology can be thought as a balance between microstructural and phenomenological formulations. The Landau phase transition formalism, provides a phenomenological description that allows to introduce the remodelling effect exerted by vimentin without the details associated with the microstructural origin of the steric interaction, which would demand a more detailed description. Nevertheless, our aim is to  provide a useful model to improve the characterisation of this kind of experiments helping with the definition of better metrics based on the complete description of the nonlinear elasticity inherent to the mechanical response. 

As a future work we expect to develop experimental and theoretical studies of this composite combined networks with the aim to better characterise the role of the phase transitions controlled by the ratios of concentrations, as described above with the aim to predict the susceptibility of the emergent network to alterations of the concentrations. This kind of studies will provide a very relevant ability to predict mechanical properties for these sorts of synthetic networks, cells and tissues.

\section*{Acknowledgments}
We thanks Prof. Eugene Terentjev from the University of Cambridge for his valuable feedback. 


\end{document}